# 13 Modeling the Impact of Dilution on the Microbial Degradation of Dispersed Oil in Marine Environments

*Vicente I. Fernandez, Roman Stocker, and Gabriel Juarez*

**CONTENTS**

## 13.1  INTRODUCTION

Crude oil undergoes natural weathering processes that change its chemical composition and physical properties upon entering the water column (Boehm et al., 1982; National Research Council, 2003; Payne and McNabb, 1984; Tarr et al., 2016). Examples of environmental weathering processes include evaporation, dissolution, dispersion, photo-oxidation, microbial degradation, emulsification, sedimentation, and burial. The timescale associated with each process can range from hours to years and the extent to which any weathering process dominates depends on the location of the oil in the water column. Near the sea surface, oil is susceptible to evaporation and photo-oxidation, which occur over hours to days, whereas deeper in the water column, oil is more susceptible to microbial degradation and sedimentation, which occur over months to years (National Research Council, 2003). Compared to traditional mechanical measures, such as skimming or burning, which typically recover less than 20% of spilled oil (Lessard and DeMarco, 2000; Wadsworth, 1995), microbial degradation is a crucial process in the weathering of unrecovered immiscible hydrocarbons that persist the longest in the marine environment, such as dispersed oil droplets in the water column (Bagby et al., 2016; Head and Swannell 1999; Head et al., 2006).

Spilled oil is mechanically dispersed into oil droplets by shear stresses from wave action, turbulence, and the high-pressure jet flow from a blowout (Delvigne and Sweeney 1988; Johansen et al., 2015). The addition of chemical dispersants increases the efficiency of mechanical dispersion by reducing the oil–water interfacial tension and further reducing the overall droplet size (Li et al., 2008; Li et al., 2009; McAuliffe et al., 1980). Droplets are rendered effectively neutrally buoyant when they are below a diameter of approximately 70 µm (Atlas and Hazen, 2011), as the resuspending effect of turbulence dominates over the rising force of buoyancy. These small, neutrally buoyant oil droplets act as passive fluid particles and are transported by advection and diffusion through the water column. In particular, these small droplets do not rise to the ocean surface and so are typically not exposed to evaporation and photo-oxidation. The ultimate fate of these oil droplets and the extent and mode of their degradation by microorganisms are currently unknown (Passow and Hetland, 2016). Here we apply a new mechanistic computational model to quantify the biodegradation processes of individual droplets of crude oil in naturally diluting environments.

Existing oil spill models vary in their level of complexity, ranging from two-dimensional models that consider the trajectories of oil surface slicks to fully three-dimensional models that take into account a large number of physical, chemical, and biological processes affecting the fate of oil (Huang, 1983; Reed et al., 1999; Spaulding, 1988; Spaulding, 2017). The more processes a model includes, the more realistic yet more computationally intensive and dependent on parameterization it becomes. Therefore, practical oil spill models typically employ a Eulerian-Lagrangian framework to describe the underlying hydrodynamics of the water column at a given location (Eulerian) as well as the transport and trajectories of small parcels of water or representative oil droplets (Lagrangian), with a time step ranging from 10 to 100 minutes (Spaulding, 2017) and a grid spacing ranging from 10 to 1000 meters (French-McCay, 2004; Reed et al., 1999). At these modeling length scales, microscale structure and heterogeneity, such as droplet size or bacterial trajectories, are replaced with empirical relations obtained from field measurements or laboratory experiments. Microbial degradation, for example, is typically modelled using independent first-order degradation equations for each degradable hydrocarbon component, based on the experimentally reported half-lives of hydrocarbon compounds (French-McCay, et al. 2018; French-McCay, 2004). Without an understanding of the fundamental microscale mechanisms that influence oil biodegradation, empirical relations and fitting parameters will introduce unwanted and unnecessary error in model predictions.

In this work, we examine the impact of natural dilution of a dispersed underwater plume by ambient turbulence on the timescale of microbial oil degradation, using an extension of the Microscale Oil Degradation Model, or MODEM. Previously, we have introduced and used MODEM to show that microscale mechanisms, such as the initial encounters between bacteria and oil droplets as well as the droplet size distribution, can be a significant factor in determining the microbial degradation time of dispersed oil droplets in quiescent environments (Fernandez et al., 2019). Unlike laboratory conditions where oil, bacteria, and seawater form a closed system, turbulence in the ocean results in the continuous dilution of any locally high concentration of oil droplets. The average oil concentration of dispersed oil plumes is expected to fall below 10 PPM (parts-per-million) within a few days of a spill, whereas over the ensuing weeks to months—when microorganisms play an important role in degradation—the typical concentration is expected to be much less than 1 PPM (Lee et al., 2013; Macnaughton et al., 2003). Since the impact of encounter times is most significant when oil concentrations are low (Fernandez et al., 2019), these results indicate that estimates of microbial degradation rates based on experimental half-lives that start from high oil concentrations are likely to be strongly overestimated, even when high concentrations are present near the source of oil, such as in the case of an underwater blowout.

## 13.2 THE MICROSCALE OIL DEGRADATION MODEL—MODEM

To examine the impact of the continuous dilution that oil dispersions are subject to in aquatic environments, we use the same underlying model (MODEM) used in a recent investigation of the optimal droplet size for biological oil degradation (Fernandez et al., 2019). For additional details on the model implementation, see the appendix of this chapter and the supplemental methods in Fernandez et al. (Fernandez et al., 2019) This model focuses on the physical encounters between oil-degrading bacteria and oil droplets as a function of droplet size distribution and oil concentration and predicts the decrease in oil mass as a result of the biological degradation. There are two distinct stages considered in the degradation model: encounter and growth. For a given oil droplet, the first stage lasts until an oil-degrading bacterium first encounters and attaches to the oil droplet. The duration of this stage is inherently stochastic, but depends on the behavior of the bacteria (such as swimming and chemotaxis), the concentration of bacteria in the environment, and the size of the oil droplet. The second stage is assumed to follow deterministic dynamics as bacteria grow on the surface and consume the available carbon in the oil droplet. During this second stage, random encounters with planktonic bacteria continue to occur, but are considered negligible because in nearly all conditions the time between arrivals is much larger than the doubling time of bacteria growing on the oil droplet surface.

The stochastic encounter process of the first oil-degrading bacterium finding an oil droplet is represented using a Poisson arrival process with an arrival rate given by a diffusive encounter kernel. The arrival rate is

$$\lambda = 4\pi \left(r_{bacteria} + r_{droplet}\right)\left(D_{bacteria} + D_{droplet}\right)C_{bacteria}, \qquad (13.1)$$

where $r$, $D$, and $C$ represent the radii, diffusivity, and concentration of the respective subscripts. In this framework, inter-arrival times, and therefore the time for the first arrival, are exponentially distributed with mean $1/\lambda$. Given that the bulk concentration of bacteria, $C_{bacteria}$, changes as a function of time, we model a large number ($10^6$) of droplets with a Monte Carlo method, revising the arrival rate at each time step based on the corresponding bulk bacterial concentration.

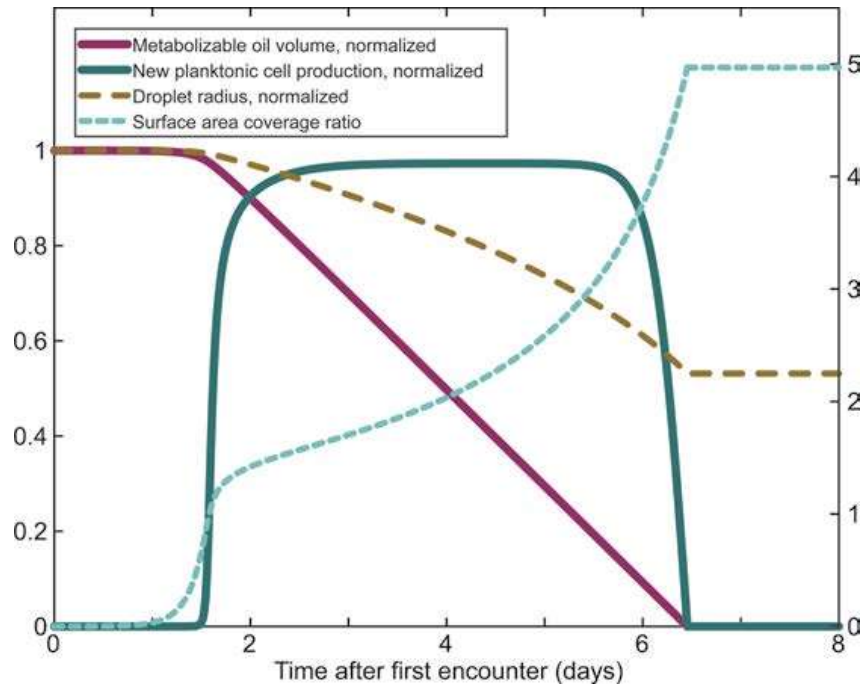

**FIGURE 13.1** Droplet degradation dynamics. Elements of the prescribed degradation dynamics for a 62 µm diameter oil droplet are depicted after the initial encounter with an oil degrading bacterium ($t = 0$). The metabolizable oil volume and the oil droplet radius are normalized by their starting values (left axis). The rate of production of new planktonic cells from the droplet is normalized by the maximum production rate (left axis). The surface area coverage ratio (see text) is not normalized (right axis).

Once the first oil-degrading bacterium has attached to an oil droplet, the degradation of the oil droplet and the growth of bacteria follow a prescribed trajectory, such as that shown in Figure 13.1 for the case of a 62 µm diameter droplet and biological parameters according to Table 13.1. Initially, bacteria attached to the droplet surface are able to grow unconstrained by space, carbon, or other nutrients. This results in exponential growth on the surface of the oil droplet, with all new bacteria remaining attached to the oil droplet.

As the surface of the oil droplet is progressively covered in bacteria, it becomes harder for all the bacteria to remain on the surface when individuals divide. The model assumes that there is a maximum number of bacteria that a droplet can host on its surface. Once this capacity is reached, all subsequent growth creates new planktonic bacteria (which are able to colonize other droplets) and the number of bacteria on the original droplet remains constant and equal to the maximum. Based on experimental observations of oil-degrading bacteria at the oil–water interface (unpublished), the maximum capacity is somewhat larger than the number of bacteria it would take to cover the original spherical surface of the droplet (Juarez and Stocker, 2019). This may be due to biofilm formation or other bacterial interactions that allow bacteria to remain attached. This biofilm is also the foundation for why we model the number of bacteria to be constant after it reaches capacity, even though the radius (and surface area) of the droplet decreases (and thus the number of bacteria

**TABLE 13.1**

**Model Parameters: Default Values of the Model Parameters unless Otherwise Stated**

| Symbol | Description | Value | Units | Reference |
|---|---|---|---|---|
| $r_{droplet}$ | Droplet radius | 31 | µm | Fernandez et al., 2019 |
| $C_o$ | Initial oil concentration | $10^{-3} - 10^4$ | PPM | |
| $D_{iso}$ | Isopycnal diffusivity | 0.07 | m²/s | Ledwell et al., 1998 |
| $D_{dia}$ | Diapycnal diffusivity | $10^{-5}$ | m²/s | Ledwell et al., 1998 |
| $T$ | Temperature | 283 | K | |
| $\eta$ | Dynamic viscosity of seawater | $1.03 \times 10^{-3}$ | Pa s | |
| $\rho$ | Oil density | 850 | kg/m³ | Daling et al., 1990; Reddy et al., 2012 |
| | Percent carbon in oil | 85 | - | Reddy et al., 2012; Stewart et al., 1993 |
| $C_{bacteria}$ | Concentration of oil-degrading bacteria | $10^3$ (initial) | cells/mL | Hazen et al., 2010 |
| $r_{bacteria}$ | Bacterial radius | 0.5 | µm | |
| $m_b$ | Carbon per bacterial cell | $100 \times 10^{-15}$ | gC | Loferer-Krößbacher et al., 1998; Vrede et al., 2002 |
| $\Gamma$ | Bacterial growth efficiency | 0.5 | - | |
| $\tau_d$ | Bacterial doubling time | $10^4$ | s | Juarez and Stocker, 2019 |

fitting on the surface would in principle decrease as well). The saturation at the maximum allowed number of attached cells is implemented as a smooth transition based on the surface area coverage ratio (see Figure 13.1, see Fernandez et al., 2019).

A consequence of the fixed number of attached bacteria once capacity is reached is that the mass of metabolizable oil decreases linearly with time during this final portion of the degradation process: a constant number of bacteria grow at a constant rate, resulting in a steady utilization of the available carbon. This proceeds until all the metabolizable carbon (assumed to be 85% of the initial oil volume (Stewart et al., 1993)) is used. Note that a loss of 85% of the volume would result in a reduction of only approximately 50% in the diameter of the droplet (see Figure 13.1). The limit to the number of bacteria that can attach to the droplet surface is responsible for the relationship between the time it takes for a droplet to degrade (post colonization) and the surface area to volume ratio, which is determined by the size of the droplet. Smaller droplets have a larger surface area to volume ratio and consequently the maximum degradation rate for a given oil volume is significantly higher. This relationship has been recognized and partially motivates the use of dispersants (Lessard and DeMarco, 2000)..

We previously used MODEM to demonstrate that for sufficiently small droplets, encounters with oil-degrading bacteria are the limiting step in the degradation, in contrast to the surface-area constrained degradation after colonization (Fernandez et al., 2019). This results in an optimal oil droplet size for biodegradation (approximately 100 µm in diameter). Although biological and environmental parameters alter the precise value of the optimal droplet size, the existence of a non-zero minimum value for the optimal droplet size was found to be robust.

During the saturated stage of degradation, the bacteria on the droplet continue to grow but new bacteria disperse into the environment. This provides a mechanism for droplets that are degraded early to affect the degradation time of remaining nearby droplets by reducing the time it takes for them to be encountered by oil-degrading bacteria (because bacteria that disperse into the environment cause an increase in $C_{bacteria}$). The increase in the concentration of planktonic oil-degrading bacteria reduces the contribution of the encounter stage to the overall degradation time, leading to a smaller optimal droplet size.

However, the practical impact of the bacteria that are released into the environment from degrading droplets depends on the concentration of oil droplets. At low oil concentrations, the fixed number of new planktonic bacteria that result from the biodegradation of an oil droplet will not meaningfully alter the background concentration $C_{bacteria}$. At the other extreme, at high oil concentrations, a single droplet may increase the local concentration of bacteria to the point of significantly decreasing the time to encounter of nearby droplets, developing into a cascade of increasing bacterial concentration and leading to a large fraction (if not all) of the nearby droplets being colonized in quick succession.

The general results for biodegradation when droplet–bacteria encounters are included hold for monodispersed as well as polydispersed oil droplets. For polydispersions, there are additional interactions between large and small droplets (Fernandez et al., 2019). However, it was found that a monodispersion that maintains the same overall surface area to volume ratio as a polydispersion is a good approximation to the polydispersion in terms of degradation time. For the following discussion of biodegradation under continuous dilution, a monodispersion of oil droplets with a 62 µm diameter will be used, representing an approximation of a theoretical log-normal droplet size distribution with a 34 µm mean diameter (Aman et al., 2015).

## 13.3  DILUTION IN AQUATIC ENVIRONMENTS

In aquatic environments, volumes of water with higher oil droplet concentrations will naturally mix over time with the surrounding oil-free water, resulting in dilution. At the scales of mixing in natural aquatic environments, a dispersion of small oil droplets can be treated as a dissolved solute for the purposes of characterizing its dilution. Thus, to predict the dilution of an oil droplet dispersion one can harness results from field studies that have tracked artificial tracers spreading over months in the oceans (Ledwell and Watson, 1991; Ledwell et al., 1998) or characterized mixing from hydrodynamic measurements (Hibiya et al., 2006; Polzin et al., 1997), as well as numerical modeling results (Simmons et al., 2004).

While droplets will spread even in completely still water with a diffusivity governed by Brownian motion, most suspended oil droplets are sufficiently large (>10 µm) that this contribution to dilution is minimal. For oil droplets 10 µm in diameter and larger, the Brownian diffusivity given by the Stokes-Einstein equation (Kiorboe, 2008) is less than approximately $2 \times 10^{-14}$ m$^2$/s, and an individual oil droplet would diffuse less than a millimeter in a day. For objects the size of oil droplets (typically, 10 – 500 µm diameter), mixing in aquatic environments is driven primarily by the ambient flow and in particular by turbulence. The initial outcome of turbulence is the stirring of a cloud into a complex geometry of thin filaments. As the scale of the filaments decreases, regular diffusion (which acts over short distances) increases in importance and smooths out heterogeneity, resulting in mixing. The overall process, "turbulent diffusion," typically behaves similarly to molecular diffusion, but with significantly higher diffusion coefficients. Away from the surface of a natural water body, any density stratification of the water suppresses mixing in the vertical direction. As a result, mixing occurs at different rates along the water density gradient (diapycnal) and perpendicular to it (isopycnal). Factors such as the dimensions of the water body and roughness of the bottom surface (Polzin et al., 1997) can also have large impacts on the rates of mixing. While dilution is a ubiquitous process affecting locally elevated concentrations of any compound in aquatic environments, its intensity thus depends strongly on the specifics of the environment.

## 13.4 DILUTION IN THE MICROSCALE OIL DEGRADATION MODEL

Although the dilution of a cloud of oil droplets by turbulence largely behaves diffusively, with regions of high concentration gradients mixing more strongly than regions with lower gradients (such as the center of an oil droplet plume), we will use some simplifications to implement dilution in conjunction with the encounter–growth oil degradation model. The primary simplification is that the model accounting for dilution will consider a single representative oil concentration instead of a range of oil concentrations that smoothly decrease from the center of the plume to the background level. For an instantaneous point source release, the spreading of a compound is given by:

$$C(x,y,z,t) = \frac{M}{(4\pi t)^{3/2} (D_x D_y D_z)^{1/2}} e^{\left(-\frac{x^2}{4D_x t} - \frac{y^2}{4D_y t} - \frac{z^2}{4D_z t}\right)}, \quad (13.2)$$

where $C$ is the concentration as a function of three-dimensional space and of time, $D_j$ is the turbulent diffusivity in the $j$ direction, $M$ is the total mass of oil injected at time zero at the origin, and $t$ is time. The solution to the 3D diffusion equation is shown in Figure 13.2. The scaling term of the equation $C_{max}(t) = M/(4\pi t)^{3/2} (D_x D_y D_z)^{1/2}$ gives the maximum concentration in space as the point source spreads. Our approximation for the dilution–degradation model will assume that, at any time $t$, the entire plume has the concentration $C_{max}(t)$ and a corresponding ellipsoidal volume that, in the absence of degradation, would conserve the initial mass of oil (see Figure 13.2, gray lines). The volume is ellipsoidal as opposed to spherical because the diffusivities are not necessarily equal in all directions. This approximation captures the critical elements of dilution, specifically the temporal dynamics of the reduction of the oil concentration, but significantly simplifies the implementation in MODEM by having only two environmental states at any given time, namely (i) water without oil and (ii) water with a single concentration of oil, $C_{max}(t)$.

In addition to tracking a single representative oil concentration, we will neglect the dependence of the diffusivity on the length scales of the oil plume (Okubo, 1971). Due to the multiple scales of motion that characterize turbulence in aquatic environments, turbulent diffusivity increases as a plume grows in size, as larger and larger eddies contribute to stirring and mixing. Here, we will neglect this size-dependence and instead use a uniform isopycnal diffusivity ($D_x = D_y = D_{iso} = 0.07$ m²/s) and a uniform diapycnal diffusivity ($D_z = D_{dia} = 10^{-5}$ m²/s) that are both independent of the size of the oil droplet cloud. The specific values are taken from a tracer release experiment performed at 300 m depth in the North Atlantic in 1992 (Ledwell et al., 1998).

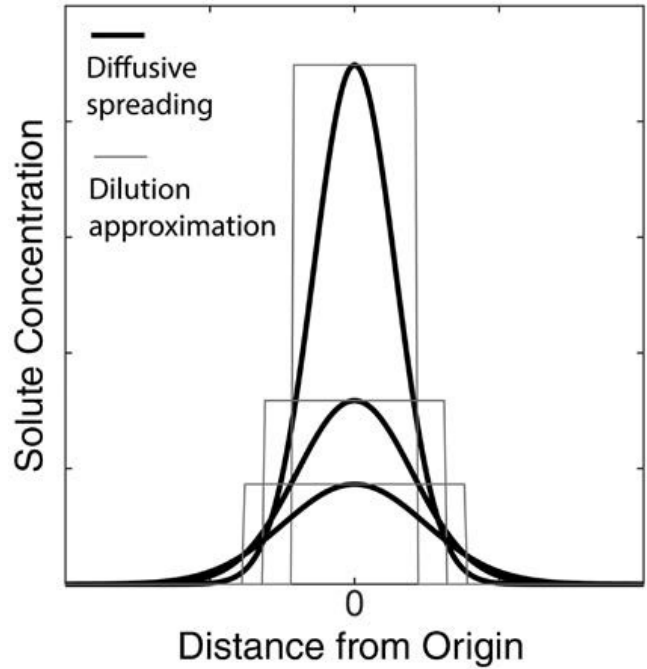

**FIGURE 13.2** Simplified dilution model, approximating diffusive spreading while retaining simplicity in the implementation of the encounter–growth model. The black lines illustrate the solution to an instantaneous point source (Eq. 13.2) starting at the origin and spreading diffusively. The gray lines illustrate the simplified dilution model, in which the concentration inside the spreading dispersion of oil droplets is uniform and equal to the maximum of the exact solution. In the simplified model, the width of the spreading distribution (gray lines) is set such that (in the absence of degradation) the total mass of oil is conserved as dilution proceeds.

The solution for the diffusion of an instantaneous point source (Eq. 13.2) results in a very rapid initial dilution, but this rapid dilution occurring in the first minutes would not be an adequate model for the oil dispersion application being considered. Although a seep in the ocean floor or a leak in a pipe may be very small and reminiscent of a point source, the process that creates the oil droplet dispersion will likely also spread the oil over a volume larger than what can be considered a point source. This initial condition is implemented here by offsetting the time by an amount $t_0$, $C_{max}(t) = C_{max}(t' + t_0)$, so that the size of the spreading point source at $t' = 0$ matches the desired initial oil plume size. As a result of this temporal offset, the volumetric dilution of oil concentration is significantly slower than $t^{-3/2}$ for $t < t_0$ (see Figure 13.3, black line).

In this numerical study, we focus on the dilution of a cloud of oil droplets that is generated due to a constant influx of oil at depth. We will assume a consistent ambient water current that advects the oil droplet dispersion, generating a plume that increases in width as the distance from the source increases. In such a plume, dilution primarily occurs in the two dimensions perpendicular to the direction of the ambient current. Along the axis of the ambient current there is a much smaller gradient in the oil concentration and therefore mixing does not have a large dilutive impact. Therefore, we can model dilution in a plume from the spreading of a cross section of the plume perpendicular to the ambient current (see Figure. 13.3, gray line). Since mixing occurs in two dimensions only, the dilution scales as $t^{-1}$ rather than as $t^{-3/2}$.

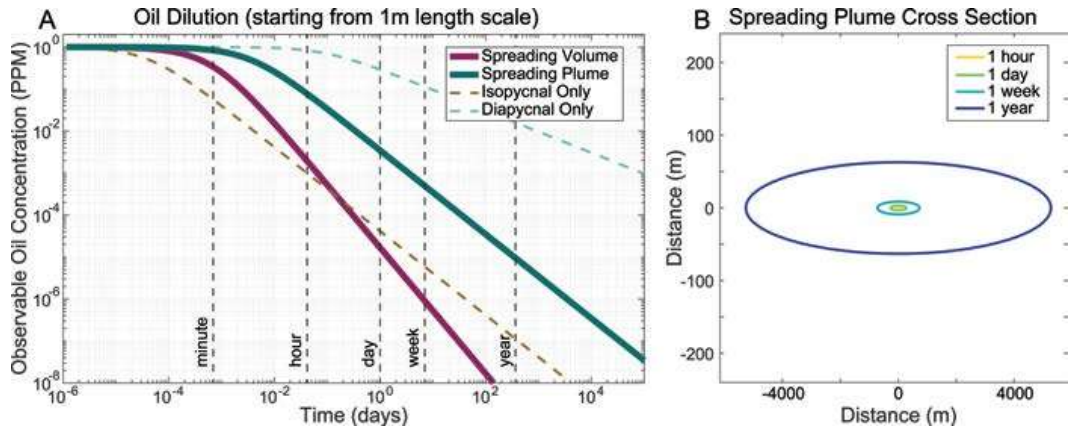

**FIGURE 13.3** Dilution of concentrated oil dispersions in the ocean. (A) The concentration of oil (in PPM = parts per million) is plotted over time for different mixing conditions. In all cases, the isopycnal and diapycnal diffusivities are 0.07 m²/s and $10^{-5}$ m²/s, respectively. The volume dilution (solid black line) initially has the volume of a 1 meter diameter sphere. The two-dimensional dilution (solid gray line) of a vertical plume cross section begins with the area of a 1 meter diameter disk, as does the horizontal two-dimensional isopycnal case (black dashed line). The one-dimensional diapycnal-only diffusion (gray dashed line) begins from a 1 meter length (B). The spreading of a plume cross section, corresponding to the solid gray line in (A). Note the different axis scaling.

The modeled oil plume cross section will have a constant width during the cross-sectional spreading. This width does not affect the simulation since it is a constant multiplicative factor when calculating the volume of the spreading plume cross section. Since we are modeling the stochastic encounter of individual droplets by directly simulating a number of droplets, the cross-sectional width is chosen for each initial oil concentration to keep the number of simulated droplets fixed at $10^6$.

As the oil droplet cloud spreads, new water from the surroundings is entrained into the oil droplet plume. Given that the encounter–growth numerical model is tracking the degradation state of individual oil droplets using a Monte Carlo approach, the increased fluid volume does not directly affect the simulated degradation of the droplets. By summing the total amount of remaining oil volume at any moment and evaluating the total volume that corresponds to the plume cross section at that time, the oil concentration can be straightforwardly calculated.

However, dilution can have a strong indirect impact on the oil degradation by altering the rate at which droplets are colonized. The new entrained water has not recently had an available oil substrate for bacteria, and therefore is assumed to have oil-degrading bacteria at the background concentration. As a result, any increases in the local concentration of planktonic oil-degrading bacteria inside the oil plume due to shedding of cells from droplets colonized to capacity is also being diluted alongside the effect of dilution on the oil droplet concentration. The concentration of oil-degrading bacteria in the diluting oil plume changes over time due to both contributions of new cells from colonized oil droplets and dilution. The impact of dilution on the concentration of bacteria was modeled by directly adding the appropriate number of new bacteria from the new volume mixed into the dispersion (having background bacterial concentration) and renormalizing by the total plume section volume:

$$N_{t+1} = N_t - \frac{\Delta t}{\kappa}\left(N_t - C_{bacteria,0}\ V_t\right) + G_t \Delta t + C_{bacteria,0}\left(V_{t+1} - V_t\right), \qquad (13.3)$$

where

$$V(t) = h\ 4\pi(t+t_0)\sqrt{D_{iso}D_{dia}}, \qquad (13.4)$$

and $N_t$ is the number of bacteria in the plume slice volume at time $t$, $C_{bacteria,0}$ is the undisturbed background concentration of bacteria, $\Delta t$ is the time step, $\kappa$ is the time for the free-swimming bacterial population to turn over (based on growth and loss rates), $G_t$ is the total rate of production of new planktonic bacteria due to oil degradation, $h$ is the width of the cross-sectional slice, and $t_0$ is the temporal offset for the appropriate starting size. The second term in Equation 13.3 represents the net predation that occurs on the bacterial population and effectively returns the population to the base concentration level in the absence of other elements stimulating growth. The third term quantifies the number of new bacteria produced by all the colonized droplets in the plume section. Since the degradation of a droplet after colonization follows a prescribed path, $G_t$ depends only on how long each simulated droplet has been colonized. Equation 13.4 describes the volume of the simplified dilution model (see Figure 13.2, gray lines) for a two-dimensional vertical cross section of the plume. This volume function is evaluated at the appropriate time steps for Equation 13.3.

With the updated concentration of bacteria within the oil droplet plume $(C_t = N_t/V_t)$, the encounter rate at time $t$ is calculated as in the original microscale oil degradation model (Eq. 13.1) as

$$\lambda_t = 4\pi\left(r_{bacteria} + r_{droplet}\right)\left(D_{bacteria} + D_{droplet}\right)\frac{N_t}{V_t}. \qquad (13.5)$$

Note that, for encounters, we remain interested in the relative diffusivity between the bacteria and the oil droplets as the process that leads to their encounters. This relative motion occurs at scales much smaller than the turbulent mixing that drives the oil dispersion dilution, which originates from velocity gradients (shear) on length scales as small as the Kolmogorov scale. While shear from turbulence or other fluid motion can in principle enhance encounter rates by producing relative motion between the bacteria and the droplet, the effect is small for most realistic

conditions(Karp-Boss et al., 1996; Kiorboe, 2008). For the encounter rate, therefore, the diffusivities in Equation 13.5 are the Brownian diffusivities of the bacteria and droplets, respectively.

With these additions, significantly affecting the concentration of bacteria, the microscale oil degradation model is here extended to include continuous dilution, such as that affecting deep-water oil droplet dispersions. The same approach could be used to model dilution near the ocean surface after dispersants and wave action disperse an oil slick; however, that case would be more appropriately modeled as a one-dimensional dilution process with a diffusivity that depends on different processes to those considered here. In the following, we shall examine how dilution in an oil droplet plume impacts the biodegradation of the oil, particularly in comparison to cases without dilution.

## 13.5  IMPACT OF DILUTION ON OIL DEGRADATION AND APPARENT OIL DEGRADATION

When we compare the temporal dynamics of oil degradation with and without dilution (see Figure 13.4), there is a large shift in the oil concentration needed to sustain a heightened bacterial presence in the ambient fluid and thus shorten the average encounter (and degradation) time. The necessary starting oil concentration is higher in the presence of dilution by three orders of magnitude, approximately $10^{-1.2}$ PPM with dilution and $10^{2.4}$ PPM without dilution. Despite the large shift in the dependence on the starting oil concentration caused by dilution, the two limiting cases of very low and very high starting oil concentration remain unchanged in the presence and absence of dilution. The dynamics at intermediate oil concentrations are qualitatively similar as well, with an initially elevated degradation rate (see the sharp decline at early times in Figure 13.4) that transitions to the exponential decay. The key effect of dilution is that larger oil concentrations are needed to sustain an increase in the background concentration of bacteria such that degradation rates are increased. Otherwise, the dilution of the oil droplet plume also dilutes any increase in bacterial concentration and returns it to the background levels.

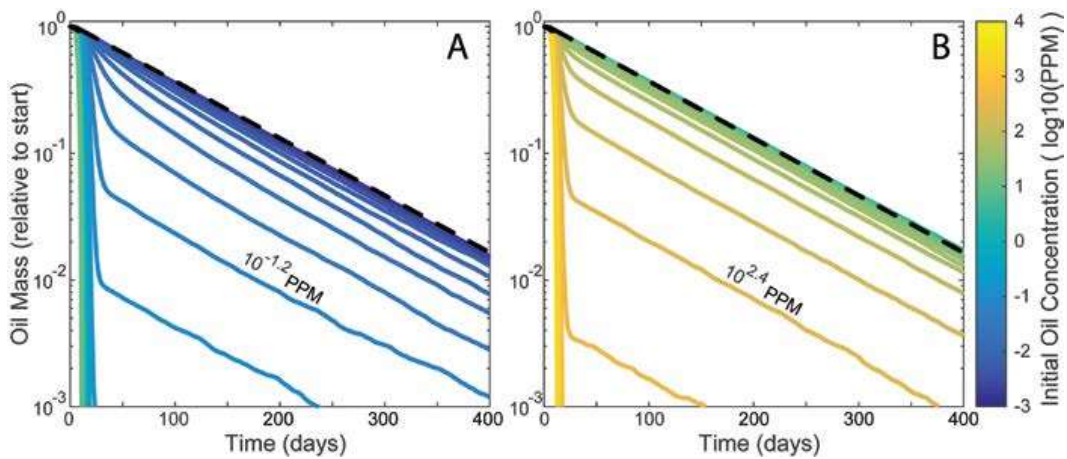

**FIGURE 13.4** Comparison of oil degradation with and without dilution. The total oil mass, normalized by its initial value, is shown as oil is degraded over time with (B) and without (A) dilution. Both plots use the same color map to indicate the starting oil droplet concentration (in PPM). In both cases, degradation occurs exponentially for low starting oil concentrations, and the limit for low starting oil concentration is the same (dashed black line). The two cases also show similar dynamics for very high starting oil concentrations. However, for intermediate concentrations, the transition to enhanced degradation rates occurs at much higher oil concentration in the presence of dilution.

In the absence of dilution and any changes to the background concentration of oil degrading bacteria, the degradation of a population of oil droplets progresses exponentially, as illustrated here for the case of 62 μm diameter droplets (see Figure 13.4, black dashed lines). The rate of the exponential decay is determined by the rate of arrivals in the encounter process, which in turn is dependent on the instantaneous concentration of bacteria in the ambient fluid. This rate is also the low oil concentration limit, since at very low oil concentrations the additional bacteria that are released from degrading oil droplets will not impact the overall bacteria concentration in the plume. Dilution further lowers the concentration of oil over time, but it does not affect degradation in this limiting case. As a result, for starting concentrations of oil below $10^{-2}$ PPM with dilution and below $10^{1.6}$ PPM without dilution, oil degradation closely follows the same exponential limit.

At very high oil concentrations, the background concentration of bacteria will quickly increase after the first oil droplet is colonized and begins to add bacteria to the environment. Shortly afterwards, all the droplets are colonized and have begun to be degraded, leading to very rapid biodegradation that is constrained by the rate of bacterial degradation and the droplet surface area. The mechanics of this limiting case are also unaffected by dilution, which constantly reduces the bacterial population in proportion to its concentration similar to an additional mortality term, but which is overcome by sufficiently high net growth rates.

While dilution does not qualitatively change the degradation dynamics in the biophysical MODEM model, the increase in the starting oil concentration needed (for the mixing rates considered here) to trigger accelerated degradation has significant implications for bioremediation strategies. As visible in Figure 13.4, there is a very substantial difference in the degradation time when it is exponential and driven by encounters (black dashed lines) in comparison to the rapid (near-linear) degradation that occurs when colonization encounters are enhanced by a large local bacterial concentration. In the example of 62 μm diameter droplets, degradation of 90% of the oil occurs in <20 days when the starting oil concentration is high enough, compared to >400 days when starting at a low concentration. In oil spills, any interventions that shift the degradation dynamics away from the encounter-limited case will accelerate degradation, but this is harder to achieve under conditions with strong dilution.

Dilution affects oil droplet degradation by altering the conditions early in the degradation process. Without dilution, there is an initial lag as the first droplets are encountered and go through the exponential growth phase. At that point, the cascade of increasing bacterial concentration in the environment may or may not begin, depending on the concentration of oil. During the initial lag time, very little if any oil is being degraded. However, with dilution, the initial lag is long enough for the local oil concentration in the plume to drop by two orders of magnitude (see Figure 13.5).

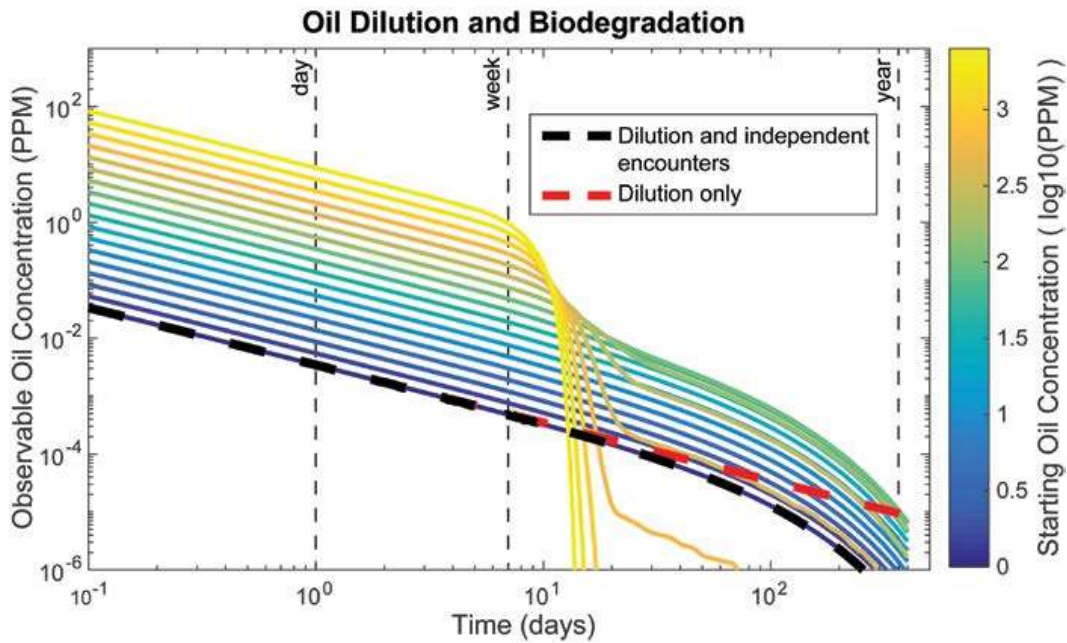

**FIGURE 13.5** Evolution of oil concentration with degradation and dilution. The oil concentration in a spreading cross section of a plume is shown over time as it is reduced by both biodegradation and dilution. The color of each curve corresponds to the starting oil concentration, indicated by the color bar. Dilution alone would produce a linear degradation when plotted logarithmically (red dashed line, compare with Figure 13.3). In the absence of any change of the background bacterial concentration, referred to here as the 'independent' limit, the reduction in oil concentration begins to significantly deviate from the dilution-only case after approximately one week (black dashed line). At higher starting oil concentrations, degradation of the oil results in an increase in the background concentration of bacteria, which leads to a faster reduction in oil concentration due to the more rapid colonization of droplets.

Under the simulated conditions, significant degradation of the oil in the plume begins to be noticeable a week into the dilution. In fact, the oil concentration a week after the initial release compares favorably to the starting concentrations without dilution (see Figure 13.4A) for predicting whether encounters will dominate the degradation or not. This suggests that the concentration a week after oil is introduced into the environment determines whether bacteria grown on oil droplets can meaningfully enhance their local background concentration.

Figure 13.5 also makes it apparent that directly observing oil droplet biodegradation in the field may be difficult, and could potentially result in the misinterpretation of dilution. When the conditions allow for a large increase in ambient bacterial concentration such that degradation is enhanced (i.e., starting oil concentrations $>10^{2.4}$ PPM), the oil concentration begins to decrease quite sharply after one week and the oil is largely degraded by bacteria over the following two weeks. In contrast, when the oil concentration is not sufficiently high to directly drive an increase in ambient bacterial concentration, the difference between dilution alone and dilution coupled with microbial degradation may not be noticeable for more than one month (see Figure 13.5, dashed lines). In addition, this evidence of biodegradation would occur at oil concentrations below $10^{-2}$ PPM, which are difficult to accurately measure in the field (Joint analysis group for the Deepwater Horizon Oil, 2011). As a result, based on our model taking encounters between bacteria and droplets into account, microbial degradation of an oil droplet dispersion should be challenging to directly measure in the field even when contributing significantly to the overall oil degradation.

## 13.6  INCREASING BIODEGRADATION RATES THROUGH ENCOUNTERS

The Deepwater Horizon spill event in 2010 generated a significant oil plume at a depth of approximately 1100 meters extending approximately 35 km over several months (Camilli et al., 2010), and provides a point of reference and comparison for the conclusions of our dilution-enhanced MODEM study. In analyses and characterization following the spill, some authors reported no significant biodegradation of the Deepwater Horizon oil plume on a timescale of months (Camilli et al., 2010), whereas others found enrichment in oil-degrading bacteria in the plume and signs of reduction in oil concentrations that were tentatively attributed to biodegradation (but acknowledged dilution as a possibility) (Hazen et al., 2010).

Despite the reports of increased bacterial concentrations at the Deepwater Horizon plume, it is difficult to evaluate whether the encounter rates with oil droplets would have been meaningfully impacted. In the study reporting enrichment of bacteria in the plume, the concentration of bacteria (of all types) was found to double from $2.7 \times 10^4$ cells/mL to $5.5 \times 10^4$ cells/mL (Hazen et al., 2010). At the level of taxa, the highest enrichment was a factor of 3 (Hazen et al., 2010). Given that the encounter rate (see Equation 13.1) is directly proportional to the bacterial concentration, this indicates that the observed enrichment at most tripled the encounter rate of droplet-degrading bacteria with oil droplets in the plume. In addition, it appears that the majority of the microbial enhancement was in response to hydrocarbon gasses in the water (Valentine et al., 2010). This raises the possibility that the bacterial populations that were initially enriched were not able to degrade oil droplets. As a result, studies of the Deepwater Horizon spill leave open the possibility that oil droplets were diluted to sufficiently low concentrations before encountering oil-degrading bacteria, resulting in a prolonged environmental persistence.

Given the likelihood that natural marine conditions lead to encounter-limited degradation due to the effect of dilution, artificially enhancing the droplet encounter rates shortly after oil enters the environment would have a large impact on biodegradation rates. One possibility for increasing encounter rates is to increase oil droplet sizes, as discussed in our previous work on the optimal droplet size (Fernandez et al., 2019). This approach is robust to dilution, since it does not rely on a subsequent enhancement of the bacterial population in the ambient fluid to increase encounter rates. A further approach for reducing the encounter time would consist of increasing the concentration of oil-degrading bacteria that are able to attach to droplets. This could potentially be done by artificially seeding oil-degrading, droplet-attaching bacteria near the source of a spill or by introducing substrates favoring the growth of such bacteria. Based on the MODEM simulations, methods for maintaining higher concentrations of oil for a sufficiently long period of time (>1 week) would also aid the encounter problem, but that would involve reducing the dilution or increasing the oil concentration.

## 13.7  CONCLUSIONS

This chapter discussed a biophysical model for hydrocarbon consumption of suspended oil droplets by bacteria under conditions of rapid dilution that occur in natural environments. Regardless of whether it is through the manipulation of droplet size or by increasing the local bacteria concentration, the results of the model indicate that stimulating the initial colonization of droplets is critical for rapid biodegradation of oil droplets in the ocean. We have found in this study that dilution heavily exacerbates the lag until oil-degrading bacteria find oil droplets, because the ambient microbial enhancement is delayed by approximately a week by which point the concentration in an oil plume has fallen. Given the complexity of physical and biological processes involved in biodegradation, the importance of physical encounters between droplets and bacteria in rapidly diluting environments has not been appreciated.

While this work has examined how dilution impacts the encounter process that is a prerequisite for oil droplet degradation, we have abstracted dilution into a single process for a spreading plume well below the ocean surface. It is important to understand microscale microbial encounters with oil droplets in the context of their environments. In natural environments, the strength of the dilution varies greatly with the location, depending on characteristics such as water depth, currents, weather, and distance from the surface. The dilution process is also dependent on how the oil dispersion is created. If an oil slick is dispersed by wave action, much of the dilution will be one dimensional, whereas a brief leak that releases a localized cloud of droplets will spread in three dimensions with stronger dilution than a plume in the same environment. Further work examining this range of conditions is necessary to understand the magnitude of the impact of encounters between droplets and bacteria under realistic oil spill environments.

## 13.8  APPENDIX

The model (MODEM) used in this study represents the biodegradation of droplets through interactions with a single bacterial species representing the first colonizers of the droplets. The model simplifies the biological complexity of a hydrocarbon-degrading bacterial community to maintain the focus on the impact of the microscale physical environment. The characteristics of the modeled bacteria are described in Table 13.1. In addition, 85% of the starting oil volume is assumed to be biodegradable by these bacteria.

MODEM also assumes that the encounter process between oil droplets and bacteria is driven primarily by relative diffusion and that the impact of fluid flow is comparatively small. This is appropriate for neutrally buoyant small entities that are jointly advected by the local flow. Once colonized by the first viable bacterium, MODEM switches from the encounter stage to the growth stage that neglects subsequent encounters. This assumes that the doubling time of bacteria on a droplet is significantly shorter than the encounter time.

MODEM directly simulates individual oil droplets over time. The primary variables tracked in MODEM are the colonization times of each droplet and the abundance of bacteria in the surrounding fluid that changes over time. For this study, the volume of the simulation also changes over time due to the expanding plume cross section.

In the simulations discussed in this chapter, each of the one million droplets has the same 62 μm diameter, matching the surface area to volume ratio of a more complex distribution of droplet sizes (see Section 13.2). The large number of droplets considered was sufficient that the degradation simulations were highly repeatable, with only small fluctuations developing in the exponential tail of oil volume below 1% of the starting volume (e.g., see Figure 13.4).

At each one-hour time step, uniformly distributed random numbers are generated for each uncolonized droplet and compared to a threshold corresponding to the probability of colonization over 1 hr given the updated encounter rate for that time step. If above the threshold, a droplet is marked as colonized at that time step. This process is repeated, updating the simulation variables at each time step, until the remaining metabolizable oil fraction drops below a critical threshold (0.1% of starting oil volume).

In order to reduce computation time and repetition, the deterministic degradation trajectories for metabolizable oil volume, surface area coverage, and newly generated planktonic bacteria (see Figure 13.1) are precomputed for all droplet sizes using a significantly finer time step than for droplet encounters (30 sec) in order to accurately capture the transitions between stages. For a given time in the simulation, the contribution of each droplet to the change in ambient bacterial abundance is determined from this precomputed information and the difference in time between the current simulation time and the colonization time of each droplet. Coupled with population turnover and new cells from the changing volume (see Equation 13.3), the bacterial abundance is updated each time step. With the further addition of the deterministic update in oil plume volume based on the total time elapsed (see Equation 13.4), the encounter rates can be calculated for each time step (see Equation 13.5) and used to simulate which of the remaining bacteria-free droplets become colonized at that time.

Modeling the Impact of Dilution                                                231Daling, P.S., Brandvik, P.J., Mackay, D., Johansen, O. (1990). Characterization of crude oils for environmental purposes. *Oil Chem Pollut* 7, 199–224.

Delvigne, G.A.L., Sweeney, C.E. (1988). Natural dispersion of oil. *Oil Chem Pollut* 032, 281–310.

Fernandez, V.I., Stocker, R., Juarez, G. (2019). Optimal droplet sizes for microbial degradation of dispersed oil: tradeoffs between physical encounters and consumption by marine bacteria. *Submitted*.

French-McCay, D. (2004). Oil spill impact modeling: development and validation. *Environ Toxicol Chem* 23, 2441–2456.

French-McCay, D.P., et al. 2018. Modeling distribution, fate, and concentrations of Deepwater Horizon Oil in subsurface waters of the Gulf of Mexico. In *Oil Spill Environmental Forensics Case Studies* New York: Elsevier Inc., 683–735.

Hazen, T.C., et al. (2010). Deep-sea oil plume enriches indigenous oil-degrading bacteria. *Science* 330, 204–208.

Head, I.M., Jones, D.M., Röling, W.F.M. (2006). Marine microorganisms make a meal of oil. *Nat Rev Microbiol* 4, 173–182.

Head, I.M., Swannell, R.P.J. (1999). Bioremediation of petroleum hydrocarbon contaminants in marine habitats. *Curr Opin Biotechnol* 10, 234–239.

Hibiya, T., Nagasawa, M., Niwa, Y. (2006). Global mapping of diapycnal diffusivity in the deep ocean based on the results of expendable current profiler (XCP) surveys. *Geophys Res Lett* 33, 2–5.

Huang, J. (1983). A review of the state-of-the-art of oil spill fate/behavior models. *Int Oil Spill Conf* 313–322.

Johansen, Ø., Reed, M., Bodsberg, N.R. (2015). Natural dispersion revisited. *Mar Pollut Bull* 93, 20–26.

Joint analysis group for the Deepwater Horizon Oil, "Review of preliminary data to examine subsurface oil in the vicinity of MC252#1, May 19 to June 19, 2010," *NOAA Tech. Rep.*, no. August, pp. 1–169, 2011.

Juarez, G., Stocker, R. (2019). Buckling and deformation of droplets due to confined growth at oil-water interfaces. *Submitted*.

Karp-Boss, L., Boss, E., Jumars, P. (1996). Nutrient fluxes to planktonic osmotrophs in the presence of fluid motion. *Oceanogr Mar Biol* 34, 71–107.

Kiorboe, T. (2008). *A Mechanistic Approach to Plankton Ecology*. Princeton, New Jersey: Princeton University Press.

Ledwell, J.R., Watson, A.J. (1991). The Santa Monica Basin tracer experiment: a study of diapycnal and isopycnal mixing. *J Geophys Res* 96, 8695.

Ledwell, J.R., Watson, A.J., Law, C.S. (1998). Mixing of a tracer in the pycnocline. *J Geophys Res Ocean* 103, 21499–21529.

Lee, K., Nedwed, T., Prince, R.C., Palandro D. (2013). Lab tests on the biodegradation of chemically dispersed oil should consider the rapid dilution that occurs at sea. *Mar Pollut Bull* 73, 314–318.

Lessard, R.R., DeMarco, G. (2000). The significance of oil spill dispersants. *Spill Sci Technol Bull* 6, 59–68.

Li, Z., Lee, K., King, T., Boufadel, M.C., Venosa, A.D. (2008). Assessment of chemical dispersant effectiveness in a wave tank under regular non-breaking and breaking wave conditions. *Mar Pollut Bull* 56, 903–912.

Li, Z., Lee, K., King, T., Boufadel, M.C., Venosa, A.D. (2009). Evaluating crude oil chemical dispersion efficacy in a flow-through wave tank under regular non-breaking wave and breaking wave conditions. *Mar Pollut Bull* 58, 735–744.

Loferer-Krößbacher, M., Klima, J., Psenner, R. (1998). Determination of bacterial cell dry mass by transmission electron microscopy and densitometric image analysis. *Appl Environ Microbiol* 64, 688–694.